\documentclass[twocolumn,showpacs,showkeys,preprintnumbers,amsmath,amssymb,floatfix]{revtex4}
\usepackage{graphicx}% Include figure files
\usepackage{epsfig}

\newcommand{\bequ}{\begin{equation}}
\newcommand{\eequ}{\end{equation}}
\newcommand{\bear}{\begin{eqnarray}}
\newcommand{\eear}{\end{eqnarray}}
\newcommand{\bwd}{\begin{widetext}}
\newcommand{\ewd}{\end{widetext}}

\begin{document}

\preprint{SLAC-PUB-9629}
\preprint{May 2003}

\title{Impact of the Wiggler Coherent Synchrotron Radiation Impedance on the
Beam Instability\footnote{Work supported by the Department of Energy contract
DE-AC03-76SF00515}}

\author{Juhao Wu}
\email{jhwu@SLAC.Stanford.EDU}
\author{G.V. Stupakov}
\email{stupakov@SLAC.Stanford.EDU}
\author{T.O. Raubenheimer}
\email{tor@SLAC.Stanford.EDU}
\author{Zhirong Huang}
\email{zrh@SLAC.Stanford.EDU}
\affiliation{Stanford Linear
Accelerator Center, Stanford University, Stanford, CA 94309}

\date{\today
\\ ,,Submitted to Physical Review Special Topics---Accelerators and Beams}

\begin{abstract}
% Text of abstract
Coherent Synchrotron Radiation (CSR) can play an important role by
not only increasing the energy spread and emittance of a beam, but
also leading to a potential instability. Previous studies of the
CSR induced longitudinal instability were carried out for the CSR
impedance due to dipole magnets. However, many storage rings
include long wigglers where a large fraction of the synchrotron
radiation is emitted.  This includes high-luminosity factories
such as DAPHNE, PEP-II, KEK-B, and CESR-C as well as the damping
rings of future linear colliders.  In this paper, the instability
due to the CSR impedance from a wiggler is studied assuming a
large wiggler parameter $K$. The primary consideration is a low
frequency microwave-like instability, which arises near the pipe
cut-off frequency. Detailed results are presented on the growth
rate and threshold for the damping rings of several linear
collider designs. Finally, the optimization of the relative
fraction of damping due to the wiggler systems is discussed for
the damping rings.
\end{abstract}



\maketitle

% main text
\section{Introduction}
When an electron bunch passes the curved trajectory inside a
dipole, the coherent synchrotron radiation (CSR) induces an
impedance \cite{Mur97,Derb95,Warnock90,KYNg90}. The CSR impedance
is known to impact single pass bunch compressors where the beam
currents are extremely high \cite{BraunPRL,BraunPRST,H01,L02} but
it is also possible that CSR might cause a microwave-like beam
instability in storage rings.  A theory of such an instability in
a storage ring has been recently proposed in Ref. \cite{SH02}
where the impedance is generated by the synchrotron radiation of
the beam in the storage ring bending magnets.

A similar instability may arise due to coherent synchrotron
radiation in long wigglers.  Many storage rings have used large
wiggler systems to reduce the damping times and control the beam
emittance.  In particular, the high luminosity colliding beam
factories DAPHNE \cite{DAPHNE}, PEP-II \cite{PEPII}, KEK-B
\cite{KEKB}, CESR-C \cite{CESRC} all have wiggler systems that can
radiate significantly more power than is radiated in the arc
bending magnets. Similarly, damping ring designs for future linear
colliders rely on large wiggler systems to obtain the required
damping rates where, again, more power can be radiated in the
wiggler systems than in the arc bending magnets.

In a previous study, the wakefield and impedance in a wiggler with
large parameter $K$ has been obtained \cite{WRS02} based on the
work of Saldin {\it{et al.}} \cite{Saldin98}. In this paper, we
study the impact of the wiggler synchrotron radiation impedance on
the beam longitudinal dynamics in rings with dipoles and wigglers.
We focus our attention on the limit of a large wiggler parameter
$K$ because this is the most interesting case for many
applications.

The paper is organized as follows. In Sec. \ref{SecII}, we will
briefly review the theory of the beam instability in a ring due to
the CSR impedance \cite{SH02}. Using the synchrotron radiation
impedance of a wiggler \cite{WRS02}, we study the beam instability
problem for a ring including both dipoles and wigglers in Sec.
\ref{SecIII}. In Sec. \ref{SecIV}, we then focus on a
low-frequency microwave-like CSR instability near the pipe cut-off
frequency which is the primary limitation. Finally, in Sec.
\ref{SecV}, we discuss effect of the CSR instability on the design
optimization of damping rings for future linear colliders.
Finally, we give some discussions and conclusion in Sec.
\ref{SecVI}.

\section{Longitudinal beam instability}\label{SecII}
We consider the longitudinal dynamics of a thin coasting beam. The
beam can be described with a longitudinal distribution function
$\rho(\nu,s,z)$. The positive direction for the internal
coordinate $s$ is pointing to the direction of the motion. The
relative energy offset of a particle having energy $E$ with
respect to the nominal energy $E_0$ is expressed as
$\nu=(E-E_0)/E_0$. The position of the reference particle in the
beam line is $z=c\,t$ with $c$ equal to the speed of light in
vacuum.

The longitudinal beam instability due to the synchrotron radiation
impedance in a ring has been studied theoretically by Stupakov and
Heifets \cite{SH02}. In this section, we briefly review and
reproduce the equations we need for our study in this paper. The
reader is advised to refer to reference \cite{SH02} for the
details of the theory.

For a matched beam, one can use a 1-D Vlasov equation to
describe the evolution of the longitudinal distribution function
$\rho(\nu,s,z)$.
    \bear\label{vlasov}
    0&=&\frac{\partial \rho}{\partial z}-\eta\,\nu\,\frac{\partial \rho}
    {\partial s}
    \nonumber \\
    &-&\frac{r_0}{\gamma}\frac{\partial \rho}{\partial \nu}
    \int^{\infty}_{-\infty}d\,\nu^{\prime}\int^s_{-\infty}d\,s^{\prime
    }\,w(s-s^{\prime})\,\rho(\nu^{\prime},s^{\prime},z)\;,
    \eear
where, the slippage factor is defined as
    \bequ
    \eta=\alpha-\frac 1{\gamma^2}\;,
    \eequ
with $\alpha$ equal to the momentum compaction factor. For ultrarelativistic
beams, we have $\eta\approx\alpha$. Based on our definition of $s$, the sign
convention for $\alpha$ is then the following. If a particle with a higher
energy than the nominal energy, i.e., $\nu>0$, would go to the head of the
bunch, then the beam line provides a negative $\alpha$. According to this
convention, then in a simple bending magnet, we will have $\alpha>0$, while in
a wiggler, we have $\alpha<0$.

\begin{table*}[htb]
\renewcommand{\tabcolsep}{2pc} % enlarge column spacing
\renewcommand{\arraystretch}{1.2} % enlarge line spacing
\begin{tabular}{|l|l|l|l|} \hline
        & NLC & TESLA & ATF \\ \hline
Circumference   $C$/km & 0.3 & 17   & 0.14 \\ \hline Dipole radius
$R$/m   & 5.5   & 80 & 5.7 \\ \hline Total bending angle
$\Theta/2\pi$  & 1 & 5/3 & 1 \\ \hline Momentum compaction
$\alpha/10^{-4}$ & 2.95 & 1.2 & 19 \\ \hline Synchrotron frequency
$Q_s$/kHz & 3.5  & 0.8  & 17.4 \\ \hline Extracted X emittance
$\gamma\epsilon_{x}/10^{-6}$m& 3 & 8 & 5 \\ \hline Extracted Y
emittance $\gamma\epsilon_{y}/10^{-8}$m& 2 & 2 & 5 \\ \hline
Energy  $E$/Gev & 1.98 & 5 & 1.3 \\ \hline Energy rms spread
$\nu_0/10^{-4}$& 9.09 & 9 & 6 \\ \hline Bunch rms length
$\sigma_z$/mm   & 3.6 & 6 & 5   \\ \hline Particles in a bunch
$N_e/10^{10}$  & 0.75 & 2 & 1 \\ \hline Wiggler peak field $B_w$/T
& 2.15  & 1.5 & 1.88 \\ \hline Wiggler period $\lambda_w$/m   &
0.27  & 0.4 & 0.4 \\ \hline Wiggler total length $L_w$/m     &
46.24 & 432 & 21.2 \\ \hline Wiggler $\beta$-function
$\beta_{x,w}$/m & 1.87  & 6.67 & 6 \\ \hline Pipe radius $b$/cm  &
1.6   & 2 & 1.2   \\ \hline $F_w$ & 2.2   & 13.4  & 1.8
\\ \hline
%ISR loss    $U_w$   /MeV    & 0.5   & 15    & 0.08  \\ \hline
\hline Cut-off wavelength   $\lambda_c$/mm & 4.9 & 1.8 & 3.1 \\
\hline Threshold at cutoff (wiggler off) $N_t/10^{10}$ & 0.60 &
27.44 & 0.95 \\ \hline Threshold at cutoff (wiggler on)
$N_t/10^{10}$ & 0.52  & 24.56 & 0.76 \\ \hline Growth time at
cutoff (wigger off) $\tau/\mu s$ & 54.9 & N/A & 34.3 \\ \hline
Growth time at cutoff (wigger on) $\tau/\mu s$ & 32.9 & N/A & 6.5
\\ \hline
%Safety factor $\eta_s$ Wiggler Off & 0.8   & 13.7  & 0.76  \\ \hline
%Safety factor $\eta_s$ Wiggler On  & 0.7   & 12.2  & 0.63  \\ \hline
\end{tabular}
\caption{\label{dampingtable}Parameters and results for the NLC
main damping ring \cite{Andy02}, the TESLA damping ring
\cite{Deck9901}, and the KEK ATF prototype damping ring
\cite{ATF02}. The parameter $F_w$ is the ratio of the radiation
power emitted in the wiggler to that emitted in the arc bending
magnets defined in Eq. (\ref{I_2_Ratio}). }
\end{table*}

The distribution function $\rho$ is written as a sum of the
equilibrium distribution function and a perturbation,
$\rho=\rho_0(\nu)+\rho_1(\nu,s,z)$, with $\rho_1 \ll \rho_0$. We
look for a perturbation as \vspace{-0.2cm}
    \bequ\label{pertb}
    \rho_1=\hat{\rho}_1\,e^{-\,i\,\omega\,z/c\,+\,i\,k\,s}\;,
    \eequ
where $k$ is the wavenumber, $\omega$ the frequency. Linearizing
the 1-D Vlasov Eq. (\ref{vlasov}) leads to the dispersion relation
\cite{SH02}:
    \bequ\label{disp}
    1=\frac{i\,r_0\,c\,Z(k)}{\gamma}\int^{\infty}_{-\infty}d\nu\frac{d
    \rho_0/d\nu}{\omega+c\,k\,\eta\,\nu}\;,
    \eequ
which determines the existence of a solution as in Eq. (\ref{pertb}).
Here, $r_0 \approx 2.82\times 10^{-15}$ m is the classical electron
radius and $\gamma$ is the Lorentz factor.  In addition,
    \bequ
    Z(k)=\int^{\infty}_0d\,s\,w(s)\,e^{-\,i\,k\,s}\;,
    \eequ
is the CSR impedance and the wake Green function $w(s)$ describes
the interaction of two particles due to the synchrotron radiation
where $w(s)\neq 0$ for $s>0$ while $w(s)=0$ for $s<0$. The
positive values of $w(s)$ correspond to the energy loss and the
negative values imply the energy gain.

\begin{figure}[htbp]
\begin{center}
\hspace*{-5mm}\mbox{\epsfig{file=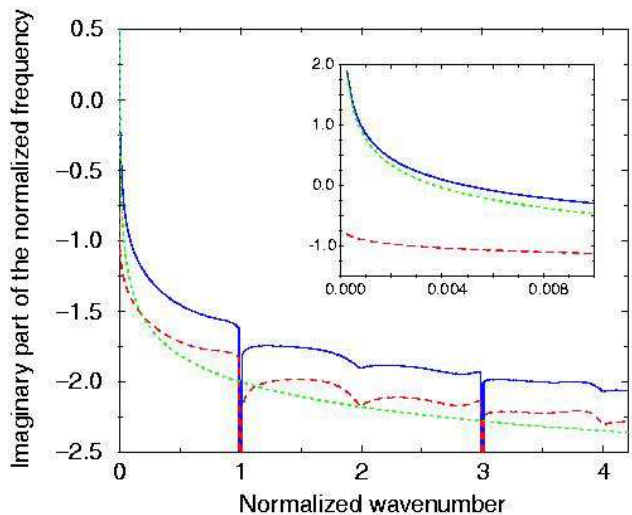, width=8.5cm}}
\end{center}
\caption{\label{nlcgrow}The imaginary part of the normalized
frequency $\Omega$ as a function of the normalized wavenumber
$k/{k_0}$ for the NLC main damping ring \cite{Andy02} where $k_0$
is the on-axis wiggler fundamental radiation wavenumber defined in
Eq. (\ref{FELFreq}). The solid curve includes the entire CSR
impedance while the dotted and dashed curves include either the
steady state dipole CSR impedance or the wiggler CSR impedance,
respectively. The inset shows a blow up of the low frequency
region where the beam is unstable. } \label{wakept}
\end{figure}

We assume that the initial energy distribution function is a
Gaussian, i.e., $\rho_0 = n_0 / (\sqrt{2\pi} \, \nu_0) \times
\exp(- \, \nu^2 / 2 \, \nu_0^2)$, where $n_0$ is the linear
density, i.e., the number of particles per unit length. In this
case, Eq. (\ref{disp}) can be rewritten as, \vspace{-0.2cm}
    \bequ\label{disprev}
    1 = - \frac{i \, Z(k) \, \Lambda} {\sqrt{2\,\pi}\,k}
    \int^{\infty}_{-\infty}d\, p\,\frac{p\,e^{-\frac{p^2}2}}
    {\Omega\pm p}\;
    \eequ
where, $\Lambda=n_0\,r_0/(|\eta|\,\gamma\,\nu_0^2)$; $\Omega =
\omega/(c\,k\, |\eta|\,\nu_0)$ and $p=\nu/\nu_0$. The upper
(lower) sign in Eq. (\ref{disprev}) refers to the case of a
positive (negative) $\eta$.  We refer the reader to Ref.
\cite{SH02} for further details.

\section{Storage Rings with Wigglers}\label{SecIII}
As mentioned, a number of storage rings utilize long damping
wigglers where a significant fraction of the energy loss per turn
is emitted. As concrete examples, we study the NLC main damping
ring \cite{Andy02}, the TESLA damping ring \cite{Deck9901}, and
the KEK ATF prototype damping ring \cite{ATF02}. Parameters are
given in Table \ref{dampingtable}.

In our paper, we use the steady state CSR impedance and assume a
distributed impedance model. For a dipole, the steady state CSR
impedance is \cite{Mur97,Derb95}
    \bequ\label{ZDipole}
    Z_D(k)=-\,i\,A\,\frac{k^{1/3}}{R^{2/3}}\;,
    \eequ
with \vspace{-0.2cm}
    \bequ
    A=3^{-1/3}\,\Gamma\left(\frac23\right)(\sqrt{3}\,i-1)\;.
    \eequ
The wiggler impedance $Z_W(k)$ is computed in Ref. \cite{WRS02}.
Hence, the total impedance is then
    \bequ\label{combimp}
    Z(k)=Z_D(k)\frac{\Theta\,R}{C}+Z_W(k)\frac{L_W}{C}\;,
    \eequ
where, $R$, $\Theta$, $L_W$ and $C$ are the dipole bending radius,
the total bending angle, the wiggler total length and the damping
ring circumference given in Table \ref{dampingtable} respectively.

To study the instability, we numerically solve the dispersion
relation Eq. (\ref{disprev}). Since we are assuming a coasting
beam model, we only consider instability wavelengths short
compared to the bunch length.

\begin{figure}[htbp]
\begin{center}
\hspace*{-5mm}\mbox{\epsfig{file=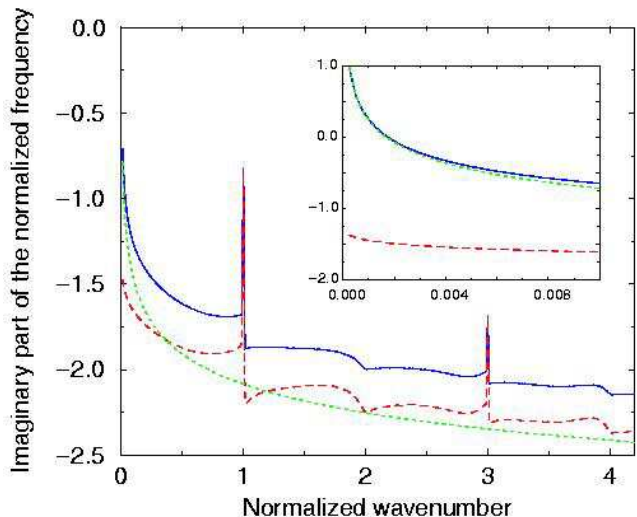, width=8.5cm}}
\end{center}
\caption{\label{nlcgrowneg}The imaginary part of the normalized
frequency $\Omega$ as a function of the normalized wavenumber
$k/{k_0}$ for the NLC main damping ring with negative momentum
compaction where $k_0$ is the on-axis wiggler fundamental
radiation wavenumber defined in Eq. (\ref{FELFreq}). The solid
curve includes the entire CSR impedance while the dotted and
dashed curves include either the steady state dipole CSR impedance
or the wiggler CSR impedance, respectively.  The inset shows a
blow up of the low frequency region where the beam is unstable. }
\label{wakept}
\end{figure}

In Fig. \ref{nlcgrow}, the imaginary part of $\Omega$ is plotted
as a function of the instability wavenumber for the NLC main
damping ring using the parameters listed in Table
\ref{dampingtable}. At low frequencies, the dipole CSR impedance
dominates while at shorter wavelengths the wiggler CSR impedance
is usually more important.  In the region where Im($\Omega$)$<0$,
the beam is stable and this is true for all regions except at the
longest wavelengths as shown in the inset.  This low frequency
instability will be discussed in the following section.

Similar calculations were made for the TESLA ``dog-bone'' damping
ring, where the total bending angle is about $\Theta=10\pi/3$, and
the KEK ATF prototype damping ring. For the design current in the
TESLA damping ring, the impedance from the dipoles and wigglers
will not drive an instability while the results for the ATF are
very similar to those of the NLC damping ring shown in Fig.
\ref{nlcgrow}.

One interesting effect can be seen at the wiggler radiation
fundamental frequency and the odd harmonics where the ring is
actually stabilized by the wiggler impedance.  This is opposite to
the single-pass behaviour through a wiggler where there is an
instability at the wiggler fundamental frequency usually refered
to as the FEL instability.  The effect will be discussed further
in Appendix \ref{Divergence} but is a direct result of the
positive sign of the momentum compaction in the ring.

To illustrate this, Fig. \ref{nlcgrowneg} is a plot for the NLC
damping ring with identical wiggler and arc parameters but the
momentum compaction is assumed to be opposite in sign.  Here, one
can see that at low frequencies the growth is slightly lower.  In
contrast, the wiggler impedance makes the system less stable at
the wiggler fundamental unlike the case with positive momentum
compaction.  Furthermore, if the magnitude of the momentum
compaction is reduced, the system will become unstable similar to
that illustrated in Appendix \ref{FEL}.  This is similar to the
FEL instability however, it is also noted in Appendix
\ref{Divergence} that our theory does not correctly treat the full
FEL instability.

\section{Low-frequency CSR instability}\label{SecIV}

As seen in Figs. \ref{nlcgrow} and \ref{nlcgrowneg}, the
instability is most important at relatively low frequency. The
longest instability wavelength that is possible is determined by
two effects: first, the instability wavelength has to be short
compared to the bunch length and, second, the vacuum chamber
causes an exponential suppression of the synchrotron radiation at
wavelengths $\lambda$ greater than the `shielding cutoff'
\cite{Warnock90} \vspace{-0.2cm}
    \bequ\label{cutoff}
    \lambda_c\leq 4\sqrt{2}b(b/R)^{1/2}\;.
    \eequ
Here, $R$ is the dipole bending radius, and $b$ is the vacuum
chamber half height. The numerical coefficient of Eq.
(\ref{cutoff}) assumes that the vacuum chamber is made up of two
infinitely wide plates. Different cross sections give different
numerical factors \cite{SK02}. Given the previous discussion, the
threshold will be the lowest at the smaller of the bunch length or
the `shielding cutoff' wavelength.

\begin{figure}[htbp]
\begin{center}
\hspace*{-5mm}\mbox{\epsfig{file=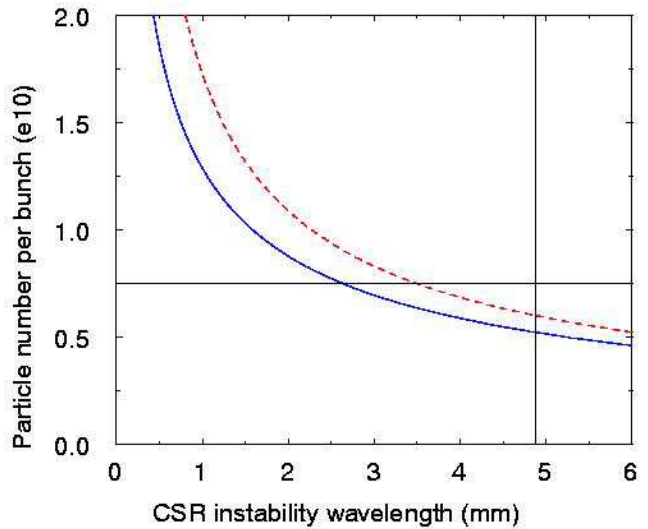, width=8.5cm}}
\end{center}
\caption{\label{nlcthreshold}The threshold particle number as a
function of the CSR wavelength for the NLC main damping ring.  The
dashed curve is the result for the dipoles only, while the solid
curve takes into account of the contributions from the dipoles and
the wigglers. The vertical straight line is the approximate cutoff
wavelength according to Eq. (\ref{cutoff}). The horizontal
straight line is the nominal number of particles per bunch:
$0.75\times 10^{10}$. }
\end{figure}
\begin{figure}[htbp]
\begin{center}
\hspace*{-5mm}\mbox{\epsfig{file=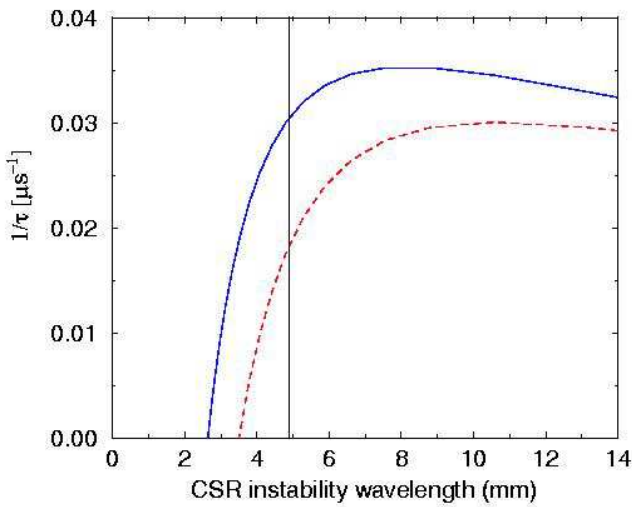, width=8.5cm}}
\end{center}
\caption{\label{nlcgrowlow}The growth rate as a function of the
CSR wavelength for the NLC main damping ring. The dashed curve is
the result for the dipoles only, while the solid curve takes into
account of the contributions from the dipoles and the wigglers.
The vertical straight line is the approximate cutoff wavelength
according to Eq. (\ref{cutoff}). }
\end{figure}

For the NLC main damping ring, we find that perturbations with
wavelengths $\lambda>3.5$ mm are not stable due to the dipole CSR
impedance alone. Adding the CSR impedance from the wiggler causes
perturbations with wavelengths $\lambda>2.6$ mm to be unstable. In
Fig. \ref{nlcthreshold}, the threshold particle number is plotted
as a function of the perturbation wavelength. Next, in Fig.
\ref{nlcgrowlow}, the growth rate, which is defined as the inverse
of the time needed for the perturbation to grow by a factor of $e$
    \bequ
    \frac 1{\tau}\equiv {\mathrm{Im}(\Omega)ck|\eta|\nu_0}\,.
    \eequ
is plotted versus the perturbation wavelength. Based on the
parameters in Table \ref{dampingtable} and Eq. (\ref{cutoff}), the
`shielding cutoff' wavelength is computed to be $\lambda_c\approx
4.9$ mm. At this cutoff wavelength, the threshold currents and
growth time are summarized in Table \ref{dampingtable}. The growth
time is significantly faster than the synchrotron period, in
agreement with the analysis for a microwave-like instability.

\begin{figure}[htbp]
\begin{center}
\hspace*{-5mm}\mbox{\epsfig{file=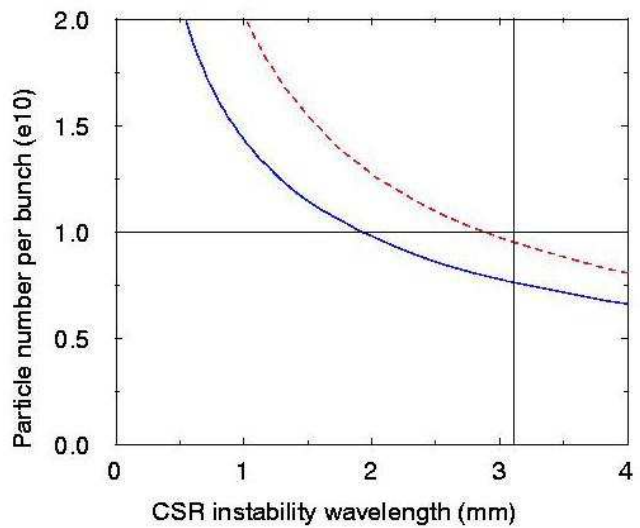, width=8.5cm}}
\end{center}
\caption{\label{kekthreshold}The threshold particle number as a
function of the CSR wavelength for the KEK ATF prototype damping
ring.  The dashed curve is the result for the dipoles only, while
the solid curve takes into account of the contributions from the
dipoles and the wigglers. The vertical straight line is the
approximate cutoff wavelength according to Eq. (\ref{cutoff}). The
horizontal straight line is the nominal number particles per
bunch: $1.0\times 10^{10}$. }
\end{figure}

It is clearly seen that the threshold current decreases as we
approach the longer wavelength perturbations, however, the growth
rate is not monotonic. This is the result of two opposite
mechanisms: one is the energy modulation growth due to the CSR
impedance, and the other is the Landau damping. For very long
wavelength perturbation, Landau damping effect is small, hence, we
could expand the denominator of the integrand in Eq. (\ref{disp})
to get \vspace{-0.2cm}
    \bequ
    \omega=\sqrt{i c^2 n_b r_0 \eta k Z(k) / \gamma},
    \eequ
which indicates that $\tau^{-1}\propto$ Im($\omega$) $\propto
\sqrt{kZ(k)}$, i.e., the growth rate will increase for a shorter
wavelength perturbation, since $Z(k)\propto k^{\varepsilon}$, with
$\varepsilon>0$. This is shown in Fig. \ref{nlcgrowlow}: when we
approach from the long wavelength perturbation to the short
wavelength perturbation, the growth rate increases.

However, this process stops when Landau damping becomes effective.
The finite energy spread in the beam will produce a phase mixing
due to the slippage and this will destroy density modulation due
to the CSR induced energy modulation. The Landau damping due to
the phase mixing is more serious for short wavelength
perturbations. This damping can be seen in the second term of Eq.
(\ref{vlasov}), or more clearly, in the denominator of the
dispersion relation in Eq. (\ref{disp}). This is demonstrated in
Fig. \ref{nlcgrowlow}, the growth rate finally decreases when we
approach very short wavelength, and eventually, the system becomes
stable.

For the KEK ATF prototype damping ring, the cut-off wavelength
would be about $\lambda_c\approx 3.1$ mm according to Eq.
(\ref{cutoff}). In Fig. \ref{kekthreshold}, the threshold particle
number is plotted as a function of the perturbation wavelength.
Taking the dipole CSR impedance alone, for the single bunch
charge, the instability sets in for perturbations with wavelengths
$\lambda>2.8$ mm. Adding the wiggler CSR impedance, the electron
beam would be unstable for perturbations with wavelengths
$\lambda>1.9$ mm. Other results are summarized in Table
\ref{dampingtable}.

It is interesting to note that in both the NLC and the ATF damping
rings roughly twice as much synchrotron radiation power is emitted
in the wiggler as in the arc dipoles.  However, the instability
threshold is not dramatically impacted by the additional radiation
in the wiggler and decreases by less than a factor of two in each
case. This arises because of the very different low frequency
behavior of the impedances.

We can use a simple scaling analysis to understand this. According
to Eq. (\ref{ZDipole}), the dipole CSR impedance scales as
$Z_D(k)\propto k^{1/3}$.  Similarly, the low frequency behavior of
the wiggler impedance is \cite{WRS02}
    \bequ\label{ZLowExact}
        Z_W(k)=\pi\,k_w\frac k{k_0}\left[1
    -\frac{2\,i}{\pi}\,\log\left(\frac k{k_0}\right)\right]\;
        \eequ
which is accurate enough for practical calculation in $k \in
[0,0.1\,k_0]$, where $k_0$ is the wiggler fundamental radiation
wavenumber: \vspace{-0.2cm}
    \bequ\label{FELFreq}
    k_0 = 2\gamma^2 k_w/ (1+K^2/2)\;
    \eequ
and the wiggler parameter $K$ is approximately $K\approx
93.4\,B_w\, \lambda_w$, with $B_w$ the peak magnetic field of the
wiggler in units of Tesla and $\lambda_w$ the period in meters.
Thus, the wiggler CSR impedance scales as Re($Z_W(k)$)$\propto k$,
and Im($Z_W(k)$)$\propto k\log(k)$, which have a weaker scaling
than $Z_D(k) \propto k^{1/3}$ when $k\rightarrow 0$, i.e., the CSR
impedance from the wiggler is weaker than that from the dipole
when we approach low frequency region.

Hence, it suggests an optimization of the damping ring design
where a larger fraction of radiation is emitted in the wigglers as
will be discussed subsequently.

\section{Damping Ring Optimization}\label{SecV}

Damping rings are used in linear colliders to attain the very
small transverse emittances that required to attain the desired
small spot sizes at the collision point.  In the damping rings,
the emission of synchrotron radiation damps the phase space volume
of the injected beams towards the equilibrium values.  An
$e$-folding reduction of the transverse phase space occurs after
the emission of synchrotron radiation energy comparable to the
beam energy.

There are usually three conflicting requirements on the damping
rings: first, the equilibrium emittances must be small which
usually requires weak bends and strong focusing; second, the
dynamic aperture must be large to accept the injected beams and
this is difficult to attain in a strong focusing storage ring;
and, third, the damping rates must be fast so that the incoming
phase space and transients can be damped before the beam are
extracted and accelerated to the collision point.

To achieve these conflicting requirements most damping ring
designs have been forced to include long damping wigglers where a
significant fraction of the energy loss per turn is emitted.  Of
course, the ring characteristics can be greatly modified by
changing the ratio of the damping due to the wiggler versus that
due to the bending magnets.  However, historically the wiggler
systems have been a concern because they are a significant source
of non-linearity and because they generate intense radiation which
may directly or indirectly impact the beam.

Other work has improved the confidence in the dynamic aperture
predictions and has shown that properly designed wigglers should
not impose a limitation \cite{marco03}.  Thus, we can further
optimize the ring designs based on quantitative evaluations of the
effect of the wiggler radiation fields.

To study the ring optimization, we use expressions derived in Ref.
\cite{ER02} for the damping ring parameters. The primary
constraint is the damping times.  Because all linear collider
designs presently only consider flat beams with
$\sigma_x\gg\sigma_y$, the tightest requirement is on the vertical
damping time which, using Eq.\ (11) of Ref. \cite{ER02}, can be
written \vspace{-0.2cm}
    \bequ\label{dampingtime}
    \tau_y\approx\frac{(2.89\times10^{12}\mathrm{kG})C}{|B_a|\gamma^2(1
    +F_w)c}\frac{|\Theta|}{2\pi}\;,
    \eequ
where $B_a$ is the arc bending field, $C$ is the ring
circumference, and $\Theta$ is the total bending angle of the
arcs. In addition, $F_w$ is a parameter equal to the ratio of the
damping due to the wiggler over the damping due to the arc bending
magnets which can be written: \vspace{-0.2cm}
    \bequ\label{I_2_Ratio}
    F_w\equiv I_{2w}/I_{2a}\;.
    \eequ
Here, $I_{2a}$ and $I_{2w}$ are the second synchrotron integrals
calculated over the arcs and the wigglers respectively, which can
be written
    \bequ\label{I_2}
    I_{2a}=\left|\frac{\Theta}{2\pi}\frac{2\pi B_a}{(B\rho)}\right|
    =\left|\frac{\Theta}R\right|;\qquad I_{2w} \approx\frac{L_w
    B_w^2}{2(B\rho)^2}\;.
    \eequ
where $(B\rho)$ is the standard energy dependent magnetic
rigidity, $(B\rho) \approx0.017\gamma$ (kG m), and $B_a$, $B_w$,
and $L_w$ are the magnetic field of the bending dipole, the peak
wiggler field, and the wiggler length.

According to Eq. (\ref{dampingtime}), we can scale the bend field as \vspace{-0.2cm}
    \bequ
    B_a\propto{1\over(1+F_w)}
    \eequ
for a given ring size and damping time and, given a maximum
wiggler field, the required wiggler length can be found from Eqs.
(\ref{I_2}). At this point, we can calculate the other ring
parameters. According to the dispersion relation in Eq.
(\ref{disprev}), we will need the momentum compaction factor
$\alpha$, the energy spread $\nu_0$, and the bunch rms length
$\sigma_z$.  Using Eqs. (37), (39) and (40) from Ref. \cite{ER02},
the scaling can be written:
    \bear
    \alpha \propto (1+F_w)^{5/3};\quad&\ &\quad
    \nu_0\propto\sqrt{{B_a+B_wF_w}\over{1+F_w}};\nonumber\\
    \sigma_z&\propto&\nu_0\sqrt{\alpha}\;.
    \eear
Examples of the scaling of these parameters versus $F_w$ can be
found in Figs. 11 and 13 of Ref. \cite{ER02} for an earlier
version of the NLC damping ring.  Using these relations, we
estimated parameters for NLC ring with $F_w=0$ and $F_w=6.5$.
These parameters are summarized in Table II along with the nominal
parameters where $F_w=2.2$.

\begin{table*}[htb]
\label{wigglerringtable}
\renewcommand{\tabcolsep}{2pc} % enlarge column spacing
\renewcommand{\arraystretch}{1.2} % enlarge line spacing
\begin{tabular}{|l|l|l|l|} \hline
$F_w$   & 0 & 2.2   & 6.5   \\ \hline
Dipole field $B_a$/T & 3.8  & 1.2   & 0.55  \\ \hline
Dipole radius $R$/m     & 1.7   & 5.5   & 12.0  \\ \hline
Wiggler length $L_w$/m & 0  & 46.24 & 66    \\ \hline
Momentum compaction $\alpha/10^{-4}$    & 0.69 & 2.95 & 14 \\ \hline
Energy rms spread $\nu_0/10^{-4}$   & 13    & 9.09 & 8.4 \\ \hline
Bunch rms length $\sigma_z$/mm  & 2.4   & 3.6   & 5.5   \\ \hline
Cutoff wavelength $\lambda_c$/mm & 8.7  & 4.9   & 3.3   \\ \hline
Threshold at cutoff $N_t/10^9$ & 2.0    & 5.2   & 31.3  \\ \hline
\end{tabular}
\caption{Parameters and results for the NLC main
damping ring when $F_w$=0, 2.2 and 6.5\;.
}
\end{table*}

\begin{figure}[htbp]
\begin{center}
\hspace*{-5mm}\mbox{\epsfig{file=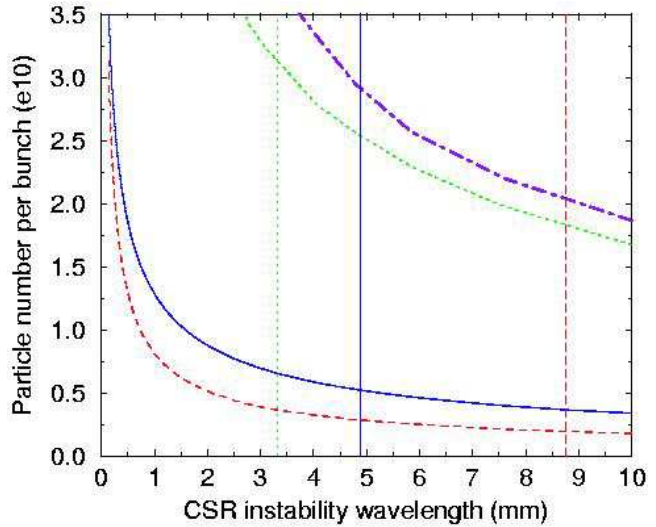, width=8.5cm}}
\end{center}
\caption{\label{fw0to10}The threshold particle number as a
function of the CSR wavelength for the NLC main damping ring.  The
dashed curve is for $F_w=0$ with the vertical dashed straight line
indicating the corresponding cutoff wavelength. Their cross point
gives the threshold at the cutoff wavelength. The solid curve and
vertical solid straight line are for the case of $F_w=2.2$, the
nominal case in Table \ref{dampingtable}. The dotted curve and
vertical dotted straight line are for the case of $F_w=6.5$. The
dot-dashed curve is for the case of $F_w=6.5$, but the impedance
is the dipole CSR impedance alone. }
\end{figure}

At this point, we can calculate the impedance.  Based on our
previous studies, we expect the CSR impedance from the wiggler to
be small compared to the CSR impedance of dipoles in the low
frequency region of interest. Hence, the threshold is essentially
determined by the dipole CSR impedance. According to the impedance
in Eqs. (\ref{combimp}) and (\ref{ZDipole}), and the relation of
$B_a$ and $F_w$ in Eq. (\ref{dampingtime}), the total impedance
should scale as
    \bequ
    Z(k)\propto Z_D(k)R \propto R^{1/3}\propto (1+F_w)^{1/3}\;.
    \eequ
at a fixed wavenumber $k$.  Clearly, this is an increasing
function with $F_w$ however, according to the dispersion relation
in Eq. (\ref{disprev}), the threshold will scale as:
    \bequ
    N_t \propto {\alpha\nu_0^2\sigma_z\over Z(k)}\propto
    (1+F_w)^{13/6}\;,
    \eequ
where it was assumed that $B_a\sim B_w$, so $\nu_0$ is roughly
constant.  Thus, it is expected that increasing $F_w$ will
increase the threshold significantly.

In addition, the cutoff wavelength $\lambda_c\propto R^{-1/2}
\propto B_a^{1/2} \propto (1+F_w)^{-1/2}$. Hence, as $F_w$
increases and $B_a$ decreases, the cutoff wavelength will
decrease. Since, the threshold is lowest at long wavelength, the
threshold will increase as $F_w$ is increased due to the shift in
the cutoff wavelength as well as the change in the ring
parameters.

The above scaling analysis has been compared to numerical
calculations for the NLC damping ring. The results are given in
Fig. \ref{fw0to10} where we plot the threshold particle number as
a function of the CSR instability wavelength for $F_w=0$ (dashes),
$2.2$ (solid) and $6.5$ (dots). In addition, the cutoff wavelength
for the three cases is indicated with vertical lines using the
same notation. Clearly, the threshold increases rapidly as $F_w$
is increased as expected.  The approximately quadratic scaling
given above is followed as $F_w$ is increased from the nominal
value of 2.2 to 6.5 but the dependence is weaker as $F_w$ is
decreased to 0 because of the change in the beam energy spread
which was assumed constant.  Based on these results a new version
of the NLC damping ring has been designed with an $F_w\approx6.5$
which has a significantly higher threshold for both this CSR
instability as well as the conventional microwave instability
driven by the vacuum chamber impedance \cite{newNLCDR}.

\section{Discussion and conclusion}\label{SecVI}
In this paper, we studied a possible beam instability due to
coherent synchrotron radiation from dipole and wiggler magnets in
a storage ring.   This instability may be important in many high
luminosity factories such as DAPHNE, PEP-II, KEK-B, and CESR-C as
well as the damping rings of future linear colliders.  We used the
results to analyze the stability of the damping ring designs and
found that a possible instability due to the CSR will exist in the
NLC damping ring design and the ATF prototype damping ring at KEK.
We then discuss optimization of the rings and the pros and cons of
increasing the damping fraction due to the wigglers.  This led to
a redesign for the NLC damping which increases the threshold to
more than four times the nominal charge per bunch.

As we discussed above, in our calculation, we used steady state
impedance for both the dipoles and wigglers. In the future, it may
be important to consider transient effects in our calculation. In
addition, the shielding cutoff of the CSR should be calculated as
part of the impedance and not treated as an ad hoc limit. This is
also important because we assumed the same shielding cutoff around
the ring but in practice it will differ in the bending magnets and
the wigglers.  Finally, the bunch length is the same order of the
instability wavelength, while in the theory, we use a coasting
beam model.  In the future, the finite bunch length should be
taken into consideration.  All of these topics are under study and
will be reported in a separate paper.

\begin{acknowledgments}
The authors thank A.W. Chao, P. Emma, S.A. Heifets, M. Ross, M.
Woodley of Stanford Linear Accelerator Center, S. Krinsky, J.B.
Murphy, J.M. Wang of National Synchrotron Light Source, Brookhaven
National Laboratory, and K.J. Kim of Advanced Photon Source,
Argonne National Laboratory for many discussions. Work was
supported by the U.S. Department of Energy under contract
DE-AC03-76SF00515.
\end{acknowledgments}
%\vspace{0.5 in}
\appendix

\section{Growth rate at the FEL frequency and the odd harmonics}
\label{Divergence}
In this appendix, we discuss the growth rates calculated at the
wiggler fundamental frequency, frequently refered to as the FEL
frequency, and at the odd harmonics.  The wiggler CSR long-range
wakefield is \cite{WRS02}
    \bwd
    \bequ\label{G_Fourier}
    G(\zeta)
    =-\frac1{2\zeta}
    +\frac1{2\zeta}\sum^{\infty}_{n=0}\left[J_n
    \left(\frac{2n+1}2\right)-J_{n+1}
    \left(\frac{2n+1}2\right)\right]^2
    \cos(4(2n+1)\zeta)\;.
    \eequ
    \ewd

To the leading order in $k$, this gives us the CSR impedance as
    \bwd
    \bear
    Z(k)
    &=&
    \pi k_w\frac k{k_0}\left\{1-\sum^{\infty}_{n=0}\left(
    \left[J_n\left(\frac{2n+1}2\right)-J_{n+1}\left(\frac{2n+1}2\right)
    \right]^2 H[k-(2n+1)k_0]\right)\right\}
    -
    2 i k_w\frac k{k_0}\log\left(\frac k{k_0}\right)
    \nonumber \\
    &+& \frac 12 i k_w \frac k{k_0}\sum^{\infty}_{n=0}\left\{\left[J_n
    \left(\frac{2n+1}2\right)-J_{n+1}\left(\frac{2n+1}2\right)\right]^2
    \log\left(\frac{[k-(2n+1)k_0]^2[k+(2n+1)k_0]^2}{k_0^4}\right)\right\}
    \;,
    \eear
    \ewd
where, $H(x)$ is the Heaviside step function, i.e., $H(x)=1$ for $x>0$,
$H(0)=1/2$, and $H(x)=0$ for $x<0$.

At the FEL frequency and the odd harmonics, the impedance has a
logarithmically divergent imaginary part, Im$(Z(k))\propto
\log(|\Delta k|/k)$, with $\Delta k\equiv k -(2n+1)k_0$, and the
real part is finite. We write $Z=R\,-\,i\,L$, where $R>0$ is the
resistance and $L>0$ is the inductance, and $L \gg R$. As a first
estimate, we neglect the resistance and write $Z=\,-\,i\,L$ only.
In the dispersion relation Eq. (\ref{disprev}), then
$|\Omega|\approx L \gg 1$. Now because of the factor
$\exp(-p^2/2)$, $\Omega$ is huge compared with $p$ in the
$p$-nonvanishing integral region. Hence the integrand could be
expanded. Keeping the first non-zero term, the integral is
simplified as
    \bequ
    \int^{\infty}_{-\infty}d\,p\,\frac{p\,e^{-\,\frac{p^2}2}}{\Omega\pm p}
    \approx\mp\sqrt{2\pi}\frac1{\Omega^2}\;,
    \eequ
for large $\Omega$ to get \vspace{-0.2cm}
    \bequ\label{ApproxGrowth}
    \Omega^2=\pm\frac{L\,\Lambda}{k}\;.
    \eequ
The upper (lower) sign in Eq. (\ref{ApproxGrowth}) refers to the
case of a positive (negative) $\eta$. For the rings which we study
in this paper, they all have $\eta>0$ and we take the upper sign.
Hence, Im$(\Omega)=0$, i.e., the growth rate will be zero and the
system is stable at the FEL frequency and all the odd harmonics.
This agrees with the conventional knowledge that a ring is stable
with a purely inductive impedance and a positive momentum
compaction or a purely capacitive impedance and a negative
momentum compaction.
\begin{figure}[htbp]
\begin{center}
\hspace*{-5mm}\mbox{\epsfig{file=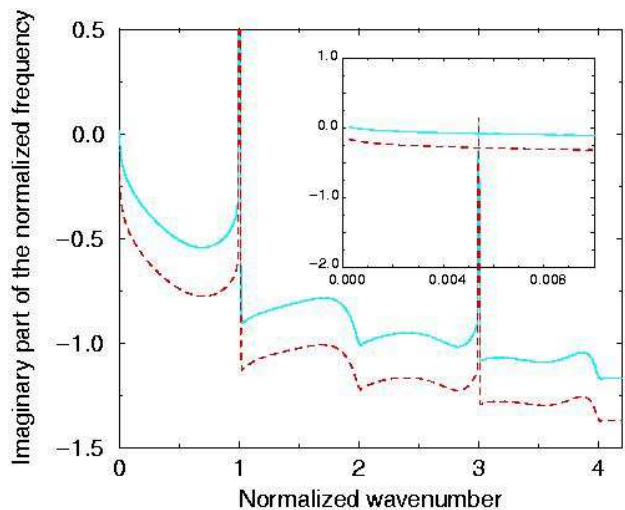, width=8.5cm}}
\end{center}
\caption{\label{dampinggrow}The imaginary part of the normalized
frequency $\Omega$ as a function of the normalized wavenumber
$k/{k_{0}}$ where $k_{0}$ is the on-axis wiggler fundamental
radiation wavenumber. The solid curve is for the wigglers in the
NLC main damping ring and the dashed curve is for the TESLA
damping ring. }
\end{figure}

\section{Single wiggler}\label{FEL}
In this appendix, we study the beam instability in a single
wiggler. From FEL theory, we would expect instability at the
wiggler fundamental radiation frequency and all the harmonics. In
addition, we need to look at a microwave-like low-frequency
instability.

In a wiggler, the slippage factor is
    \bequ\label{WigglerEta}
    \eta
    =-\frac{\left(1+\frac{K^2}2\right)}{\gamma^2}\;.
    \eequ
i.e., the slippage factor is negative, therefore, we choose the
lower ($-$) sign in Eq. (\ref{disprev}). In Fig.
\ref{dampinggrow}, we plot the imaginary part of $\Omega$ as a
function of the instability wave number for the wigglers in the
NLC and TESLA damping rings \cite{Andy02,Deck9901}. For the region
where Im($\Omega$)$<0$, the beam is stable. We find that, for the
nominal current, the system is stable within the entire
interesting region where the perturbation wavelength $\lambda<3$
cm, except at the wiggler fundamental radiation wavelength and the
third harmonic.

In fact, according to Eq. (\ref{WigglerEta}), we have $\eta<0$ and
we need take the lower sign in Eq. (\ref{ApproxGrowth}). When we
approach the fundamental and the third harmonic frequency, $L$
becomes large, and Eq. (\ref{ApproxGrowth}) gives us a large
Im($\Omega$), i.e., a large growth rate as we find in Fig.
\ref{dampinggrow}. However, according to Eq. (\ref{ApproxGrowth}),
we will have Im$(\Omega)\propto I^{1/2}$, where $I$ is the peak
current of the electron bunch; while in the FEL theory, the growth
rate scales as Im$(\Omega)\propto I^{1/3}$ \cite{BPN84}. To
correctly calculate the FEL instability from the wake impedance
approach, the electron density distribution function has to be
taken at the retarded time in the standard Vlasov equation
\cite{SK03}. This then leads to a cubic dispersion relation same
as in the FEL theory, and results in the cubic root scaling in the
growth rate: Im$(\Omega)\propto I^{1/3}$.

Nevertheless, here we further use the FEL theory to study the
instability at the FEL frequency and the third harmonic. We use
the simulation code TDA \cite{TW89} to calculate the instability
growth length at the fundamental frequency and a newly developed
code \cite{JHW02} to calculate the instability at the third
harmonic. The wiggler and beam parameters used are listed in Table
\ref{dampingtable}. In the NLC damping wiggler, the growth length
at the FEL frequency is about 50 m assuming that the phase errors
between wiggler sections is small. Fortunately, the separation
between the 2-meter long wiggler sections will further reduce the
gain. In addition, any modulation that is induced at the
fundamental wavelength of $\sim10\,\mu$m will be smeared out by
the energy spread and the momentum compaction as the beam is
transported around the remainder of the ring.  A similar
conclusion is found for the TESLA damping ring wigglers and the
instability at the FEL frequency and the odd harmonics are not
expected to be a limitation in either case.

\end{document}